\documentclass[12pt]{article}
\usepackage{graphicx}

\begin{document}

\title{Vacuum properties in the presence of quantum fluctuations of the quark
condensate.\thanks{%
Talk delivered at the workshop on \textit{Few-Quark Problems} in Bled,
Slovenia, July 8-16, 2000.}}
\author{Georges Ripka \\
Service de Physique Th\'{e}orique\\
Centre d'Etudes de Saclay\\
91191\ Gif-sur-Yvette, France\\
ripka@cea.fr}
\date{July 14, 2000}
\maketitle

\begin{abstract}
The quantum fluctuations of the quark condensate are calculated using a
regulated Nambu Jona-Lasinio model. The corresponding quantum fluctuations
of the chiral fields are compared to those which are predicted by an
''equivalent'' sigma model. They are found to be large and comparable in
size but they do not restore chiral symmetry. The restoration of chiral
symmetry is prevented by an ''exchange term'' of the pion field which does
not appear in the equivalent sigma model. A vacuum instability is found to
be dangerously close when the model is regulated with a sharp 4-momentum
cut-off.
\end{abstract}

\section{\protect\bigskip Introduction.}

This lecture discusses the modifications of vacuum properties which could
arise due to quantum fluctuations of the chiral field, more specifically,
due to the quantum fluctuations of the quark condensate. The latter is found
to be surprisingly large, the root mean square deviation of the quark
condensate attaining and exceeding 50\% of the condensate itself. We shall
discuss two distinct modifications of the vacuum: restoration of chiral
symmetry due to quantum fluctuations of the chiral field, as heralded by
Kleinert and Van den Boosche \cite{Kleinert99}, and a vacuum instability not
related to chiral symmetry restoration \cite{Ripka00}.

\section{Chiral symmetry restoration due to quantum fluctuations of the
chiral field.}

\subsection{The linear sigma model argument.}

\label{sec:linsigmod}

The physical vacuum with a spontaneously broken chiral symmetry is often
described by the linear sigma model, which, in the chiral limit $\left(
m_{\pi }=0\right) $, has a euclidean action of the form: 
\begin{equation}
I=\int d_{4}x\left( \frac{1}{2}\left( \partial _{\mu }\sigma \right) ^{2}+%
\frac{1}{2}\left( \partial _{\mu }\pi _{i}\right) ^{2}+\frac{\kappa ^{2}}{8}%
\left( \sigma ^{2}+\pi _{i}^{2}-f_{\pi }^{2}\right) ^{2}\right)
\end{equation}
Classically, we have (for translationally invariant fields): 
\begin{equation}
\sigma ^{2}+\pi _{i}^{2}=f_{\pi }^{2}
\end{equation}
and the vacuum stationary point is: 
\begin{equation}
\sigma =f_{\pi }\;\;\quad \quad \pi _{i}=0
\end{equation}

We assume that $\kappa ^{2}$ is large enough (and the $\sigma $-meson is
heavy enough) not to have to worry about the quantum fluctuations of the $%
\sigma $ field.\ So we quantize the pion field while the $\sigma $ field
remains classical. We may then say that: $\sigma ^{2}=f_{\pi
}^{2}-\left\langle \pi _{i}^{2}\right\rangle $. Classically, $\left\langle
\pi _{i}^{2}\right\rangle =0$ but the quantum fluctuations of the pion field
make $\left\langle \pi _{i}^{2}\right\rangle >0$ and therefore $\sigma
^{2}<f_{\pi }^{2}$.

Let us estimate the fluctuation $\left\langle \pi _{i}^{2}\right\rangle $ of
the pion field. A system of free pions of mass $m_{\pi }$ is described by
the partition function: 
\begin{equation}
Z=\int D\left( \pi \right) e^{-\frac{1}{2}\int d_{4}x\pi _{i}\left(
-\partial _{\mu }^{2}+m_{\pi }^{2}\right) \pi _{i}}=e^{-\frac{1}{2}tr\ln
\left( -\partial _{\mu }^{2}+m_{\pi }^{2}\right) }
\end{equation}
It follows that: 
\begin{equation}
\frac{1}{2}\int d_{4}x\left\langle \pi _{i}^{2}\left( x\right) \right\rangle
=-\frac{\partial \ln Z}{\partial m_{\pi }^{2}}=\frac{\partial }{\partial
m_{\pi }^{2}}\frac{1}{2}tr\ln \left( -\partial _{\mu }^{2}+m_{\pi
}^{2}\right) =\frac{1}{2}\Omega \left( N_{f}^{2}-1\right) \sum_{k<\Lambda }%
\frac{1}{k^{2}+m_{\pi }^{2}}  \label{pia1}
\end{equation}
where the sum is regularized using a 4-momentum cut-off and where $\Omega
=\int d_{4}x$.is the euclidean space-time volume. In the chiral limit $%
m_{\pi }=0$, we have: 
\begin{equation}
\left\langle \pi _{i}^{2}\left( x\right) \right\rangle =\frac{1}{2\Omega }%
\left( N_{f}^{2}-1\right) \sum_{k<\Lambda }\frac{1}{k^{2}}=\left(
N_{f}^{2}-1\right) \frac{\Lambda ^{2}}{16\pi ^{2}}
\end{equation}
so that: 
\begin{equation}
\sigma ^{2}=f_{\pi }^{2}-\left( N_{f}^{2}-1\right) \frac{\Lambda ^{2}}{16\pi
^{2}}  \label{condition}
\end{equation}
If we had evaluated this quantity with a 3-momentum cut-off, we would have
obtained $\left\langle \pi _{i}^{2}\right\rangle =\left( N_{f}^{2}-1\right) 
\frac{\Lambda ^{2}}{8\pi ^{2}}$. Let us pursue with a 4-momentum cut-off.\
We have: 
\begin{equation}
\frac{\left\langle \pi _{i}^{2}\right\rangle }{f_{\pi }^{2}}=\left(
N_{f}^{2}-1\right) \frac{\Lambda ^{2}}{16\pi ^{2}f_{\pi }^{2}}
\end{equation}
We deduce that chiral symmetry restoration will occur when $\sigma =0$, that
is, when $\frac{\left\langle \pi _{i}^{2}\right\rangle }{f_{\pi }^{2}}>1$: 
\begin{equation}
\frac{\left\langle \pi _{i}^{2}\right\rangle }{f_{\pi }^{2}}=\left(
N_{f}^{2}-1\right) \frac{\Lambda ^{2}}{16\pi ^{2}f_{\pi }^{2}}>1
\end{equation}
With $f_{\pi }=93\,\;MeV$ and with $N_{f}^{2}-1=3$ pions, the condition
reads: 
\begin{equation}
\Lambda ^{2}>\frac{1}{2.20\times 10^{-6}}\quad \;\;\Lambda >674\;MeV
\end{equation}
In most calculations which use the Nambu Jona-Lasinio model, this condition
is fulfilled.\ We conclude that the quantum fluctuations of the pion do
indeed restore chiral symmetry. If we had used a 3-momentum cut-off, chiral
symmetry would be restored when $\Lambda >477\;MeV$.

\subsection{The non-linear sigma model argument.}

We now argue that this is precisely what is claimed by Kleinert and Van den
Boosche \cite{Kleinert99}, although it is said in a considerably different
language. They argue as follows.\ If $\kappa ^{2}$ (and therefore the $%
\sigma $ mass) is large enough, the action can be thought of as the action
of the non-linear sigma model, which in turn can be viewed as an action with 
$N_{f}^{2}$ fields, namely $\left( \sigma ,\pi _{i}\right) $, subject to the
constraint: 
\begin{equation}
\sigma ^{2}+\pi _{i}^{2}=f_{\pi }^{2}
\end{equation}
The way to treat the non-linear sigma model is in the textbooks \cite{Zinn89}%
. We work with the action: 
\begin{equation}
I_{\lambda }\left( \sigma ,\pi \right) =\int d_{4}x\left( \frac{1}{2}\left(
\partial _{\mu }\sigma \right) ^{2}+\frac{1}{2}\left( \partial _{\mu }\pi
_{i}\right) ^{2}+\lambda \left( \sigma ^{2}+\pi _{i}^{2}-f_{\pi }^{2}\right)
\right) 
\end{equation}
in which we add a constraining parameter $\lambda $. The action is made
stationary with respect to variations of $\lambda $. We integrate out the $%
\pi $ field, to get the effective action: 
\begin{equation}
I_{\lambda }\left( \sigma \right) =\int d_{4}x\left( \frac{1}{2}\left(
\partial _{\mu }\sigma \right) ^{2}+\lambda \left( \sigma ^{2}-f_{\pi
}^{2}\right) \right) +\frac{1}{2}tr\ln \left( -\partial _{\mu }^{2}+\lambda
\right) 
\end{equation}
The action is stationary with respect to variations of $\lambda $ and $%
\sigma $ if: 
\begin{equation}
\lambda \sigma =0\;\;\;\;\;\;\sigma ^{2}=f_{\pi }^{2}-\frac{1}{2}\left(
N_{f}^{2}-1\right) \sum_{k}\frac{1}{k^{2}+\lambda }
\end{equation}
So either $\lambda =0$ and $\sigma \neq 0$, in which case we have: 
\begin{equation}
\sigma ^{2}=f_{\pi }^{2}-\frac{1}{2}\left( N_{f}^{2}-1\right) \sum_{k}\frac{1%
}{k^{2}}  \label{sol1}
\end{equation}
or $\lambda \neq 0$ and $\sigma =0$.

The condition (\ref{sol1}) is exactly the same as the condition (\ref
{condition}). Thus, the ''stifness factor'', discussed in Ref.\cite
{Kleinert99}, is nothing but a measure of $\frac{\left\langle \pi
_{i}^{2}\right\rangle }{f_{\pi }^{2}}$.

\section{Quantum fluctuations of the quark condensate calculated in the
Nambu Jona-Lasinio model.}

\label{sec:njlfluc}

We now show that the quantum fluctuations of the chiral field are indeed
large in the Nambu Jona-Lasinio model, but that chiral symmetry is far from
being restored. The regularized Nambu Jona-Lasinio model is defined in
section \ref{section:njl}.\ We begin by giving some results.

In the Nambu Jona-Lasinio\ model, the chiral field is composed of a scalar
field $S$ and $N_{f}^{2}-1$ pseudoscalar fields $P_{i}$.\ They are related
to the quark bilinears: 
\begin{equation}
S=V\left( \bar{\psi}\psi \right) \;\;\;\;P_{i}=V\left( \bar{\psi}i\gamma
_{5}\tau _{i}\psi \right) 
\end{equation}
where $V=-\frac{g^{2}}{N_{c}}$ is the 4-quark interaction strength.\ The
quark propagator in the vacuum is: 
\begin{equation}
\frac{1}{k_{\mu }\gamma _{\mu }+M_{0}r_{k}^{2}}
\end{equation}
and the model is regularized using either a sharp 4-momentum cut-off or a
soft gaussian cut-off function: 
\begin{eqnarray}
r_{k} &=&1\;if\;k^{2}<\Lambda ^{2}\;\;\;r_{k}=0\;if\;k>\Lambda \;\;\;\left(
sharp\;cut-off\right)   \label{cutoffshape} \\
\;\;\;\;\;r_{k} &=&e^{-\frac{k^{2}}{2\Lambda ^{2}}}\;\;\left(
gaussian\;regulator\right)   
\end{eqnarray}
. Let $\varphi _{0}=M_{0}$ be the strength of the scalar field in the
physical vacuum.\ We shall show results obtained with typical parameters.\
If we choose $M_{0}=300\;MeV$ and $\Lambda =750\;MeV$, then $\frac{M_{0}}{%
\Lambda }=0.4$. We then obtain $f_{\pi }=94.6\;MeV$ with a sharp cut-off and 
$f_{\pi }=92.4\;MeV$ with a gaussian cut-off (in the chiral limit). The
interaction strengths are: 
\begin{equation}
V=-9.53\;\Lambda ^{-2}\;\left( sharp\,\;cut-off\right)
\;\;\;\;\;V=-18.4\;\Lambda ^{-2}\;\left( gaussian\,\;cut-off\right) 
\label{vstrength}
\end{equation}
and the squared pseudo-scalar field has the expection value 
\begin{equation}
\left\langle P_{i}^{2}\right\rangle =V^{2}\left\langle \left( \bar{\psi}%
i\gamma _{5}\tau _{i}\psi \right) ^{2}\right\rangle 
\end{equation}
At low $q$ we identify the pion field as: 
\begin{equation}
\pi _{i}=\sqrt{Z_{\pi }}P_{i}\;\;\;\;\;\;f_{\pi }=\sqrt{Z_{\pi }}M_{0}
\end{equation}
so that, in the Nambu Jona-Lasinio model: 
\begin{equation}
\frac{\left\langle \pi _{i}^{2}\right\rangle }{f_{\pi }^{2}}=\frac{%
V^{2}\left\langle \left( \bar{\psi}i\gamma _{5}\tau _{i}\psi \right)
^{2}\right\rangle }{M_{0}^{2}}
\end{equation}
where $\left\langle \left( \bar{\psi}i\gamma _{5}\tau _{i}\psi \right)
^{2}\right\rangle $ is the pion contribution to the squared condensate.

\subsection{Results obtained for the quark condensate and for the quantum
fluctuations of the chiral field.}

\label{section:results}

Let us examine the values of the quark condensates and of the quantum
fluctuations of the chiral field calculated in the chiral limit.

\begin{itemize}
\item  The quark condensate calculated with a sharp cut off is: 
\begin{equation}
\left\langle \bar{\psi}\psi \right\rangle ^{\frac{1}{3}}=-0.352\times
\Lambda =263\;MeV\;\;\;\left( sharp\;cut-off\right) 
\end{equation}
is about 25 \% smaller when it is calculated with a soft gaussian regulator: 
\begin{equation}
\left\langle \bar{\psi}\psi \right\rangle ^{\frac{1}{2}}=-0.\allowbreak
280\times \Lambda =210\;MeV\;\;\;\left( gaussian\;regulator\right) 
\end{equation}

\item  The magnitude of the quantum fluctuations of the pion field can be
measured by the mean square deviation $\Delta ^{2}$ of the condenstate from
its classical value: 
\begin{equation}
\Delta ^{2}=\left\langle \left( \bar{\psi}\Gamma _{a}\psi \right)
^{2}\right\rangle -\left\langle \bar{\psi}\psi \right\rangle ^{2}
\end{equation}
The relative root mean square fluctuation of the condensate $\Delta $ is: 
\begin{equation}
\frac{\Delta }{\left| \left\langle \bar{\psi}\psi \right\rangle \right| }%
=0.41\quad \;\left( sharp\;cut-off\right) \;\;\;\;\;\frac{\Delta }{%
\left\langle \bar{\psi}\psi \right\rangle }=0.77\quad \;\left(
gaussian\;regulator\right) 
\end{equation}
These are surprisingly large numbers, certainly larger than $1/N_{c}$. The
linear sigma model estimate did give us a fair warning that this might occur.

\item  This feature also applies to the ratio $\frac{\left\langle \pi
_{i}^{2}\right\rangle }{f_{\pi }^{2}}$ $=\frac{V^{2}\left\langle \left( \bar{%
\psi}i\gamma _{5}\tau _{i}\psi \right) ^{2}\right\rangle }{M_{0}^{2}}$ which
was so crucial for the linear sigma model estimate of the restoration of
chiral symmetry. We find: 
\begin{equation}
\frac{\left\langle \pi _{i}^{2}\right\rangle }{f_{\pi }^{2}}%
=0.38\;\;\;\left( sharp\;cut-off\right) \;\;\;\;\frac{\left\langle \pi
_{i}^{2}\right\rangle }{f_{\pi }^{2}}=0.85\quad \;\left(
gaussian\;regulator\right) 
\end{equation}

\item  In spite of these large quantum fluctuations of the chiral field, the
quark condensates change by barely a few percent. This is shown in tables 
\ref{table:sharpcond} and \ref{table:softcond} where various contributions
to the quark condensate are given in units of $\Lambda ^{3}$. The change in
the quark condensate is much smaller than $1/N_{c}$.
\end{itemize}

\begin{table}[tbp] \centering%
%
\begin{tabular}{|c|c|c|c|}
\hline
$\left\langle \bar{\psi}\psi \right\rangle $ & $\sigma $-contribution & $\pi 
$-contribution & total \\ \hline
classical & -0.04187 & 0 & -0.04187 \\ \hline
exchange term & 0.00158 & -0.00475 & -0.00317 \\ \hline
ring graphs & 0.00014 & 0.00134 & 0.00148 \\ \hline
total contribution & -0.04015 & -0.00341 & -0.04356 \\ \hline
\end{tabular}
\caption{Various contributions to the quark condensate calculated with a
sharp 4-momentum cut-off and with $M_{0}/\Lambda =0.4$.\ The quark
condensate is expressed in units of $\Lambda ^{3}$.\label{table:sharpcond}}%
\end{table}%
%

\begin{table}[tbp] \centering%
%
\begin{tabular}{|c|c|c|c|}
\hline
$\left\langle \bar{\psi}\psi \right\rangle $ & $\sigma $-contribution & $\pi 
$-contribution & total \\ \hline
classical & -0.02178 & 0 & -0.02178 \\ \hline
exchange term & 0.00162 & -0.00486 & -0.00324 \\ \hline
ring graphs & 0.00077 & 0.00228 & 0.00305 \\ \hline
total contribution & -0.0193 & -0.00258 & -0.02197 \\ \hline
\end{tabular}
\caption{Various contributions to the quark condensate calculated with a
gaussian cut-off function and with $M_{0}/\Lambda =0.4$.\ The quark
condensate is expressed in units of $\Lambda ^{3}$.\label{table:softcond}}%
\end{table}%
%

\subsection{\protect\bigskip The effect and meaning of the exchange terms.}

The tables \ref{table:sharpcond} and \ref{table:softcond} show that, among
the $1/N_{c}$ corrections, the exchange terms dominate.\ The exchange and
ring graphs are illustrated on figures \ref{graph2} and \ref{graph13}.\ The
way in which they arise is explained in section \ref{sec:exchange}. The
exchange graphs contribute 2-3 times more than the remaining ring graphs.
Furthermore, the pion contributes about three times more to the condensate
than the sigma, so that the sigma field contributes about as much to the
exchange term as any one of the pions. However, the exchange term in the
pion channel enhances the quark condensate instead of reducing it. As a
result of this there is a very strong cancellation between the exchange
terms and the ring graphs.\ This is why the sigma and pion loops contribute
so little to the quark condensate.\ They increase the condensate by 4\% when
a sharp cut-off is used, and by 1\% when a gaussian regulator is used. This
is about ten times less than $1/N_{c}$.

The ring graphs reduce the condensate (in absolute value) in both the sigma
and pion channels. This can be expected.\ Indeed, the ring graphs promote
quarks from the Dirac sea negative energy orbits (which contribute negative
values to the condensate) to the positive energy orbits (which contribute
positive values to the condensate). The net result is a positive
contribution to the condensate which reduces the negative classical value.

What then is the meaning of the exchange terms? The exchange terms have the
special feature of belonging to first order perturbation theory (see figures 
\ref{graph2} and \ref{graph13}).\ Their contribution to the energy is not
due to a modification of the Dirac sea. It is simply the exchange term
arising in the expectation value of the quark-quark interaction in the Dirac
sea.

However, the contribution of the exchange term to the quark condensate does
involve $q\bar{q}$ excitations.\ These excitations are due to a modification
of the constituent quark mass which is expressed in terms of quark-antiquark
excitations of the Dirac sea. The exchange term is modifying (increasing in
fact) the constituent quark mass and therefore the value of $f_{\pi }$.

These results suggest that, in order to reduce the Nambu Jona-Lasinio model
to an equivalent sigma model, it might be better to include the exchange
term in the constituent quark mass, which is another way of saying that, in
spite of the $1/N_{c}$ counting rule, it may be better to do Hartree-Fock
theory than Hartree theory. The exchange (Fock) term should be included in
the gap equation. The direct (Hartree) term is, of course, included in the
classical bosonized action.

In the equivalent sigma model, $f_{\pi }$ is proportional to the constituent
quark mass. Failure to notice that that the constituent quark mass is
altered by the exchange term is what lead Kleinert and Van den Boosche to
conclude erronously in Ref.\cite{Kleinert99} that chiral symmetry would be
restored in the Nambu Jona-Lasinio model. They were right however in
expecting large quantum fluctuations of the quark condensate.

\section{The regularized Nambu Jona-Lasinio model.}

\label{section:njl}

The condensates quoted in section \ref{section:results} were calculated with
a regularized Nambu Jona-Lasinio model which is defined by the euclidean
action:

\begin{equation}
I_{m}\left( q,\bar{q}\right) =\int d_{4}x\;\left[ \bar{q}\left( -i\partial
_{\mu }\gamma _{\mu }\right) q+m\bar{\psi}\psi -\left( \frac{g^{2}}{2N_{c}}%
+j\right) \left( \bar{\psi}\Gamma _{a}\psi \right) ^{2}\right] 
\label{nonlocact}
\end{equation}
The euclidean Dirac matrices are $\gamma _{\mu }=\gamma ^{\mu }=\left(
i\beta ,\vec{\gamma}\right) $. The matrices $\Gamma _{a}=\left( 1,i\gamma
_{5}\vec{\tau}\right) $ are defined in terms of the $N_{f}^{2}-1$ generators 
$\vec{\tau}$ of flavor rotations. Results are given for $N_{f}=2$ flavors.
The coupling constant $\frac{g^{2}}{N_{c}}$ is taken to be inversely
proportional to $N_{c}$ in order to reproduce the $N_{c}$ counting rules.
The current quark mass $m$ is introduced as a source term used to calculate
the regularized quark condensate $\left\langle \bar{\psi}\psi \right\rangle $%
.\ We have also introduced a source term $\frac{1}{2}j\left( \bar{\psi}%
\Gamma _{a}\psi \right) ^{2}$ which is used to calculate the squared quark
condensate $\left\langle \left( \bar{\psi}\Gamma _{a}\psi \right)
^{2}\right\rangle $.

The quark field is $q\left( x\right) $ and the $\psi \left( x\right) $
fields are \emph{delocalized} quark fields, which are defined in terms of a 
\emph{regulator} $r$ as follows: 
\begin{equation}
\psi \left( x\right) =\int d_{4}x\;\left\langle x\left| r\right|
y\right\rangle \;q\left( y\right) 
\end{equation}
The regulator $r$ is diagonal in $k$-space: $\left\langle k\left| r\right|
k^{\prime }\right\rangle =\delta _{kk^{\prime }}r\left( k\right) $ and its
explicit form in given in Eq.(\ref{cutoffshape}).\ The use of a sharp
cut-off function is tantamount to the calculation of Feynman graphs in which
the quark propagators are cut off at a 4-momentum $\Lambda $ - a most usual
practise. The regulator $r$, introduced by the delocalized fields, makes all
the Feynman graphs converge. A regularization of this type results when
quarks propagate in a vacuum described by in the instanton liquid model of
the QCD\ (see Ref.\cite{Diakonov96} and further references therein). A Nambu
Jona-Lasinio model regulated in this manner with a gaussian regulator was
first used in Ref.\cite{Birse95}, and further elaborated and applied in both
the meson and the soliton sectors \cite{Birse95},\cite{Birse98},\cite
{Ripka98},\cite{Broniowski99}, \cite{Broniowski00},\cite{Ripka00}.\ Its
properties are also discussed in \cite{Ripka97}.

With one exception.\ In this work, as in Ref.\cite{Ripka00}, the regulator
multiplies the current quark mass. The introduction of the regulator in the
current quark mass term $m\bar{\psi}\psi =m\bar{q}r^{2}q$ requires some
explanation. The current quark mass $m$ is used as a source term to
calculate the quark condensate $\left\langle \bar{\psi}\psi \right\rangle $
which, admittedly, would be finite (by reason of symmetry) even in the
absence of a regulator - and, indeed, values of quark condensates are
usually calculated with an unregularized source term in the Nambu
Jona-Lasinio action. However, when we calculate the \emph{fluctuation } $%
\left\langle \left( \bar{\psi}\Gamma _{a}\psi \right) ^{2}\right\rangle
-\left\langle \bar{\psi}\psi \right\rangle ^{2}$ of the quark condensate,
the expectation value $\left\langle \left( \bar{\psi}\Gamma _{a}\psi \right)
^{2}\right\rangle $ diverges. It would be inconsistent and difficult to
interpret the fluctuation $\left\langle \left( \bar{\psi}\Gamma _{a}\psi
\right) ^{2}\right\rangle -\left\langle \bar{\psi}\psi \right\rangle ^{2}$
if $\left\langle \bar{\psi}\psi \right\rangle $ were evaluated using a bare
source term and $\left\langle \left( \bar{\psi}\Gamma _{a}\psi \right)
^{2}\right\rangle $ using a regulator. When a regularized source term $m\bar{%
\psi}\psi =m\bar{q}r^{2}q$ is used, the current quark mass $m$ can no longer
be identified with the current quark mass term appearing in the QCD
lagrangian. Of course, when a sharp cut-off is used, it makes no difference
if the current quark mass term is multiplied by the regulator or not.\ We
have seen in section \ref{section:results} that the leading order
contribution to the quark condensate $\left\langle \bar{\psi}\psi
\right\rangle ^{1/3}$ diminishes by only 20\% when the sharp cut-off is
replaced by a gaussian regulator. (This statement may be misleading because
when the sharp cut-off is replaced by a gaussian regulator, the interaction
strength $V$ is also modified so as to fit $f_{\pi }$. If we use a gaussian
regulator, the quark condensate calculated with a regulated source term $%
mr^{2}$ is $\left\langle \bar{\psi}\psi \right\rangle =-0.0218\;\Lambda ^{3}$
whereas the quark condensate calculated with a bare source term $m$ is $%
\left\langle \bar{\psi}\psi \right\rangle =-0.0505\;\Lambda ^{3}$.)

The way in which the current quark mass of the QCD lagrangian appears in the
low energy effective theory is model dependent and it has been studied in
some detail in Ref.\cite{Musakhanov99} within the instanton liquid model of
the QCD vacuum \cite{Shuryak82},\cite{Diakonov86},\cite{Diakonov96}.

An equivalent bosonized form of the Nambu Jona-Lasinio action (\ref
{nonlocact}) is$:$%
\begin{equation}
I_{j,m}\left( \varphi \right) =-Tr\ln \left( -i\partial _{\mu }\gamma _{\mu
}+r\varphi _{a}\Gamma _{a}r\right) -\frac{1}{2}\left( \varphi -m\right)
\left( V-j\right) ^{-1}\left( \varphi -m\right)   \label{ij}
\end{equation}
The first term is the quark loop expressed in terms of the chiral field $%
\varphi $, which is a chiral 4-vector $\varphi _{a}=\left( S,P_{i}\right) $
so that $\varphi _{a}\Gamma _{a}=S+i\gamma _{5}\tau _{i}P_{i}$. In the
second term, the chiral 4-vector $m_{a}\equiv \left( m,0,0,0\right) $ is the
current quark mass and $V$ is the local interaction: 
\begin{equation}
\left\langle xa\left| V\right| yb\right\rangle =-\frac{g^{2}}{N_{c}}\delta
_{ab}\delta \left( x-y\right) 
\end{equation}
The partition function of the system, in the presence of the sources $j$ and 
$m$ is given by the expression: 
\begin{equation}
e^{-W\left( j,m\right) }=\int D\left( \varphi \right) e^{-I_{j,m}\left(
\varphi \right) -\frac{1}{2}tr\ln \left( V-j\right) }  \label{wj}
\end{equation}
The quark condensate $\left\langle \bar{\psi}\psi \right\rangle $ and the
squared quark condensates $\left\langle \left( \bar{\psi}\Gamma _{a}\psi
\right) ^{2}\right\rangle $ can be calculated from the partition function $%
W\left( j,m\right) $ using the expressions: 
\begin{equation}
\left\langle \bar{\psi}\psi \right\rangle =\frac{1}{\Omega }\frac{\partial
W\left( j,m\right) }{\partial m}\;\;\;\;\;\frac{1}{2}\left\langle \left( 
\bar{\psi}\Gamma _{a}\psi \right) ^{2}\right\rangle =-\frac{1}{\Omega }\frac{%
\partial W\left( j,m\right) }{\partial j}  \label{condensates}
\end{equation}
where $\Omega $ is the space-time volume $\int d_{4}x=\Omega $.

The stationary point $\varphi _{a}=\left( M,0,0,0\right) $ of the action is
given by the gap equation: 
\begin{equation}
\left( V-j\right) ^{-1}=-4N_{c}N_{f}\frac{M}{M-m}g_{M}  \label{gap}
\end{equation}
This equation relates the constituent quark mass $M$ to the interaction
strength $V-j$.

\subsection{The exchange and ring contributions.}

\label{sec:exchange}

The second order expansion of the action $I_{jm}\left( \varphi \right) $
around the stationary point reads: 
\begin{equation}
I_{jm}\left( \varphi \right) =I_{jm}\left( M\right) +\frac{1}{2}\varphi
\left( \Pi +\left( V-j\right) ^{-1}\right) \varphi
\end{equation}
where $I_{jm}\left( M\right) $ is the action calculated at the stationary
point $\varphi =\left( M,0,0,0\right) $ and where $\Pi $ is the polarization
function (often referred to as the Lindhardt function): 
\begin{equation}
\left\langle xa\left| \Pi \right| yb\right\rangle =-\frac{\delta }{\delta
\varphi _{a}\left( x\right) \delta \varphi _{b}\left( y\right) }Tr\ln \left(
-i\partial _{\mu }\gamma _{\mu }+r\varphi _{a}\Gamma _{a}r\right)
\end{equation}
Substituting this expansion into the partition function (\ref{wj}), we can
calculate the partition function $W\left( j,m\right) $ using gaussian
integration with the result: 
\begin{equation}
W\left( j,m\right) =I_{jm}\left( M\right) +\frac{1}{2}tr\ln \left( 1-\Pi
\left( V-j\right) \right)  \label{saddle}
\end{equation}

\begin{figure}[h]
\centerline{\resizebox*{6.0762in}{1.6016in}{\includegraphics{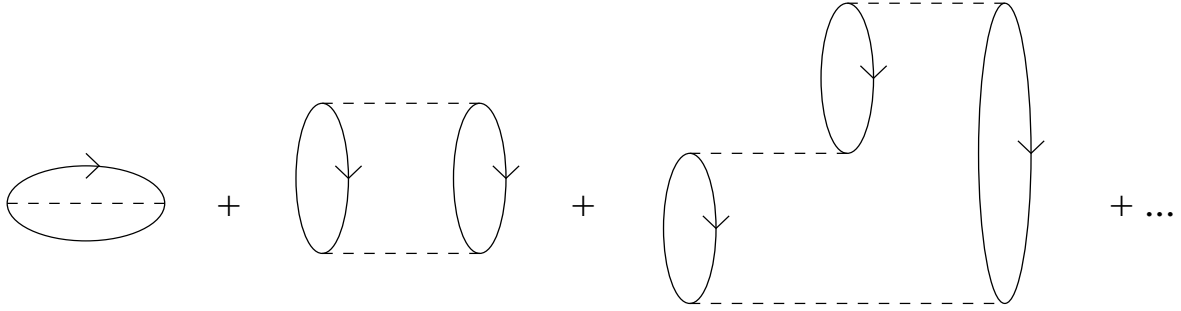}}}
\caption{The Feynman graphs
which represent the meson loop contribution to the partition function. 
The first graph is the exchange
graph and the remaining graphs are the ring graphs.}
\label{graph2}
\end{figure}

\begin{figure}[h]
\centerline{\resizebox*{12cm}{4cm}{\includegraphics{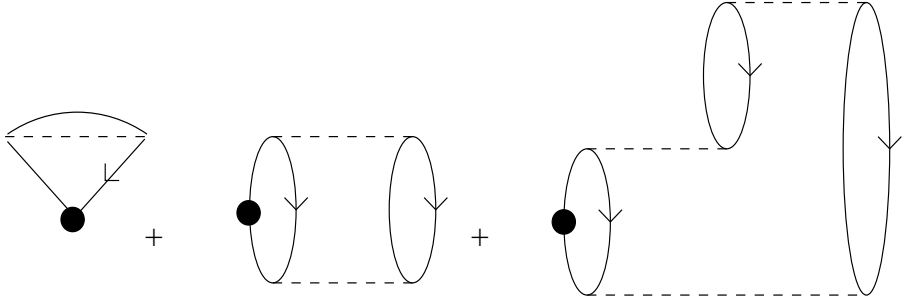}}}
\caption{The contribution to the quark
condensate of the Feynman graphs shown on figure \ref{graph2}.\ The black
blob represents the operator $\bar{\protect\psi}\protect\psi $. The first
graph (which is the dominating contribution) is the contribution of the
exchange term.\ It represents $q\bar{q}$ excitations which describe a change
in mass of the Dirac sea quarks.\ This exchange graph would not appear in a
Hartree-Fock approximation, which would include the exchange graph in the
gap equation.}
\label{graph13}
\end{figure}

The first term of the action (\ref{saddle}) is what we refer to as the
''classical'' action. The values labelled ''classical'' in the tables
displayed in section \ref{section:results} are obtained by calculating the
condensates (\ref{condensates}) while retaining only the term $I_{jm}$ in
the partition function (\ref{saddle}). The logarithm in (\ref{saddle}) is
what we refer to as the loop contribution. The expansion of the logarithm
expresses the loop contribution in terms of the Feynman graphs shown on
figure \ref{graph2}.

The first term of the loop expansion is what we call the ''exchange term'',
also referred to as the Fock term:\footnote{%
The direct (Hartree) term is included in the ''classical'' action $I_{j,m}$.}%
: 
\begin{equation}
W_{exch}=-\frac{1}{2}tr\Pi \left( V-j\right)  \label{exch1}
\end{equation}
The remaining terms are what we call the ring graphs.

stop here

It is simple to show that the inverse meson propagators are given by: 
\begin{equation}
K^{-1}=\Pi +\left( V-j\right) ^{-1}
\end{equation}

\begin{figure}[h]
\centerline{\resizebox*{5.1673in}{4.3785in}{\includegraphics{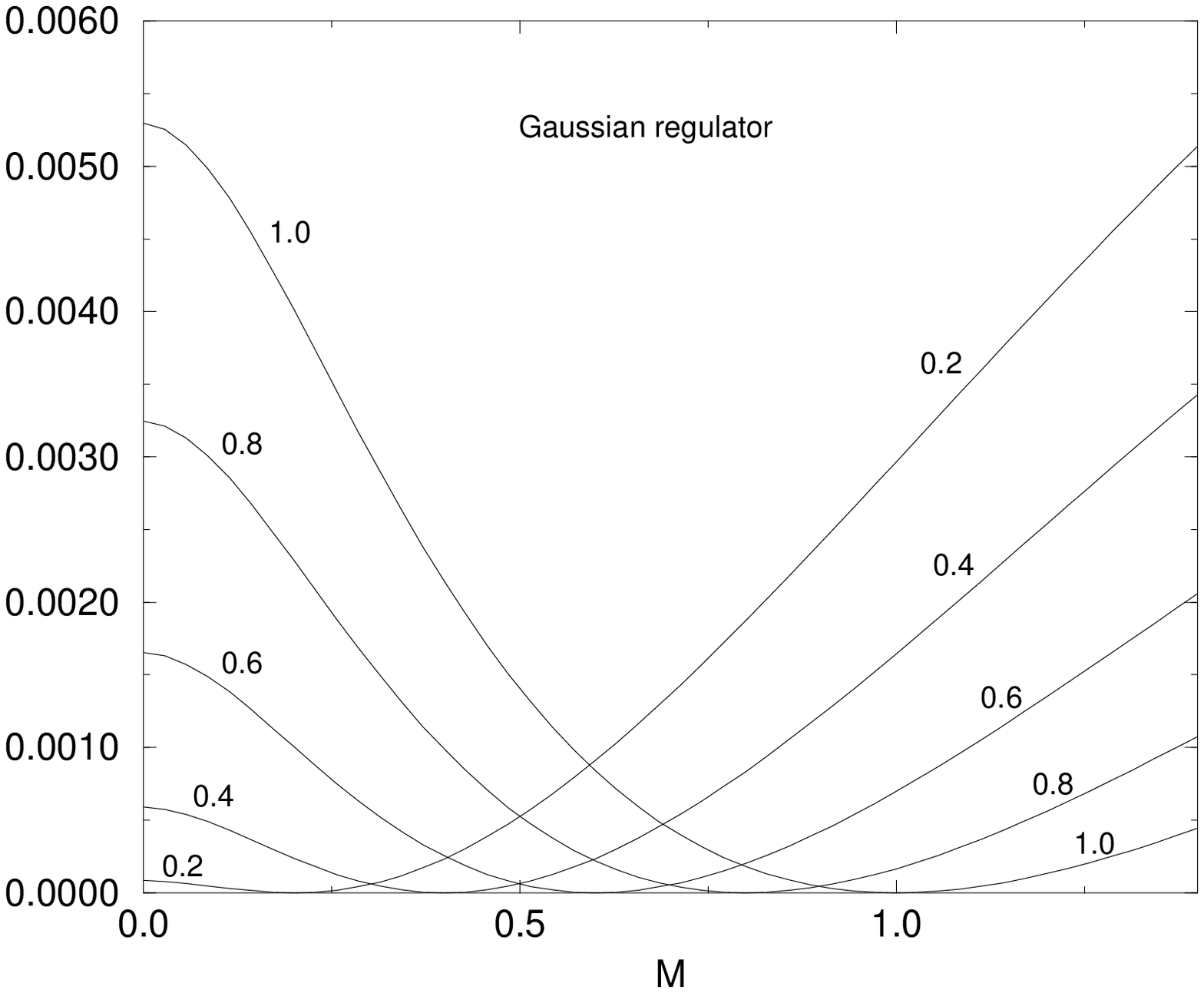}}}
\caption{The effective potential plotted
against $M$ when a soft gaussian cut-off function is used. The potential is
expressed units of $\Lambda ^{4}$.}
\label{epsfig52}
\end{figure}

They are diagonal in momentum and flavor space: $\left\langle qa\left|
K^{-1}\right| k^{\prime }q\right\rangle =\delta _{ab}\delta _{kk^{\prime
}}K_{a}\left( q\right) $ and a straighforward calculation yields the
following explicit expressions for the $S$ (sigma) and $P$ (pion) inverse
propagators:

\begin{eqnarray}
K_{S}^{-1}\left( q\right)  &=&4N_{c}N_{f}\left( \frac{1}{2}%
q^{2}f_{M}^{22}\left( q\right) +M^{2}\left( f_{M}^{26}\left( q\right)
+f_{M}^{44}\left( q\right) \right) -g_{M}\left( q\right) +\frac{M}{M-m}%
g_{M}\right)   \label{kspcorrect} \\
K_{P}^{-1}\left( q\right)  &=&4N_{c}N_{f}\left( \frac{1}{2}%
q^{2}f_{M}^{22}\left( q\right) +M^{2}\left( f_{M}^{26}\left( q\right)
-f_{M}^{44}\left( q\right) \right) -g_{M}\left( q\right) +\frac{M}{M-m}%
g_{M}\right)   
\end{eqnarray}
where the loop integrals are: 
\begin{equation}
f_{M}^{np}\left( q\right) =\frac{1}{\Omega }\sum_{k}\frac{r_{k-\frac{q}{2}%
}^{n}r_{k+\frac{q}{2}}^{p}}{\left( \left( k-\frac{q}{2}\right) ^{2}+r_{k-%
\frac{q}{2}}^{4}M^{2}\right) \left( \left( k+\frac{q}{2}\right) ^{2}+r_{k+%
\frac{q}{2}}^{4}M^{2}\right) }
\end{equation}
and: 
\begin{equation}
g_{M}\left( q\right) =\frac{1}{\Omega }\sum_{k}\frac{r_{k-\frac{q}{2}}^{2}}{%
\left( k-\frac{q}{2}\right) ^{2}+r_{k-\frac{q}{2}}^{4}M^{2}}r_{k+\frac{q}{2}%
}^{2}\;\;\;\;g_{M}\equiv g_{M}\left( q=0\right) 
\end{equation}
These are the expressions which are obtained from the second order expansion
of the action (\ref{ij}) retaining the regulators from the outset and
throughout.

\begin{figure}[h]
\centerline{\resizebox*{4.9995in}{4.2099in}{\includegraphics{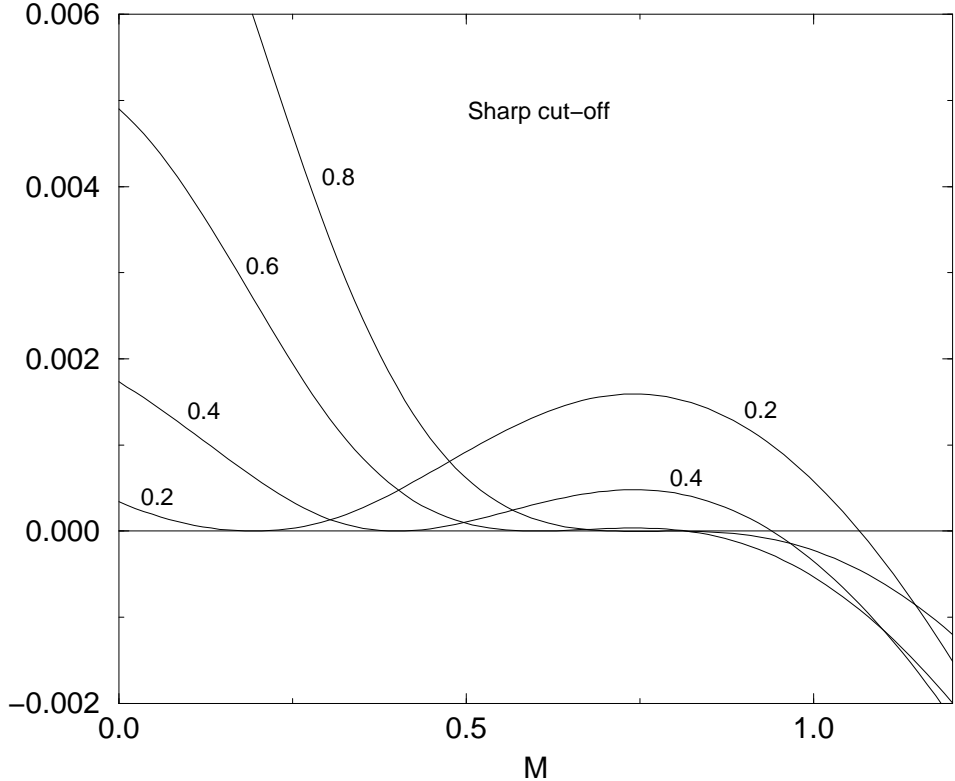}}}
\caption{The effective potential plotted
against $M$ when a sharp cut-off is used.The effective potential is
expressed in units of $\Lambda ^{4}$.}
\label{epsfig57}
\end{figure}

\begin{figure}[h]
\centerline{\resizebox*{5.1007in}{4.0992in}{\includegraphics{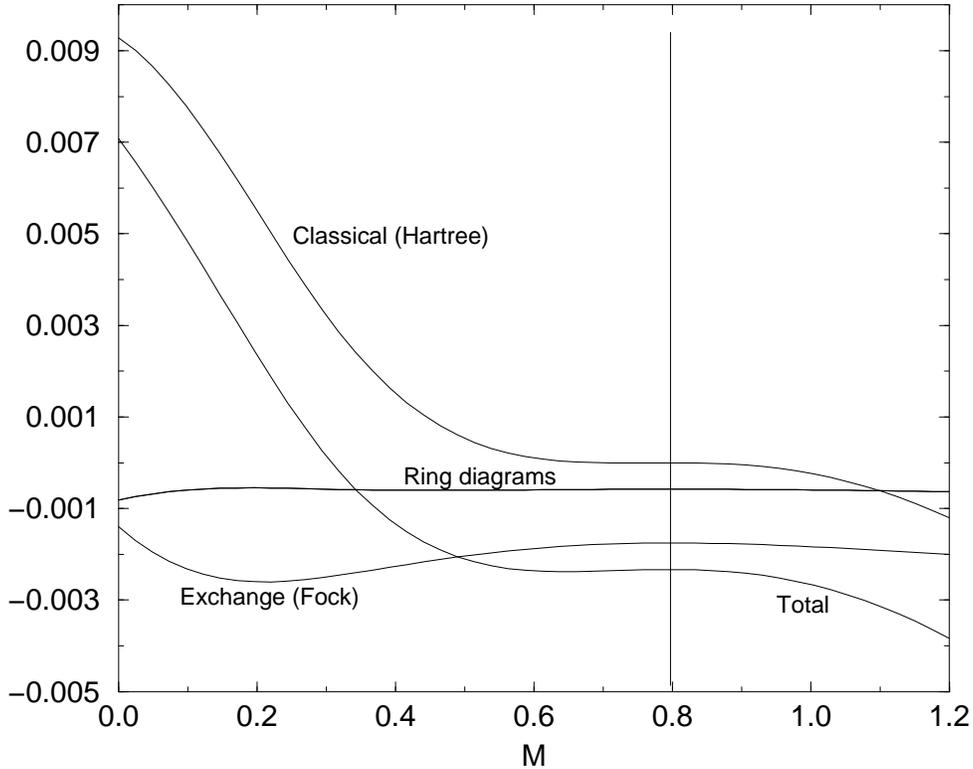}}}
\caption{Various contributions to the
effective potential calculated with a sharp cut-off and $M_{0}/\Lambda =0.8$%
. The contributions are expressed in units of $\Lambda ^{4}$.}
\label{epsfig47}
\end{figure}

\begin{figure}[h]
\centerline{\resizebox*{13.0172cm}{11.2423cm}{\includegraphics{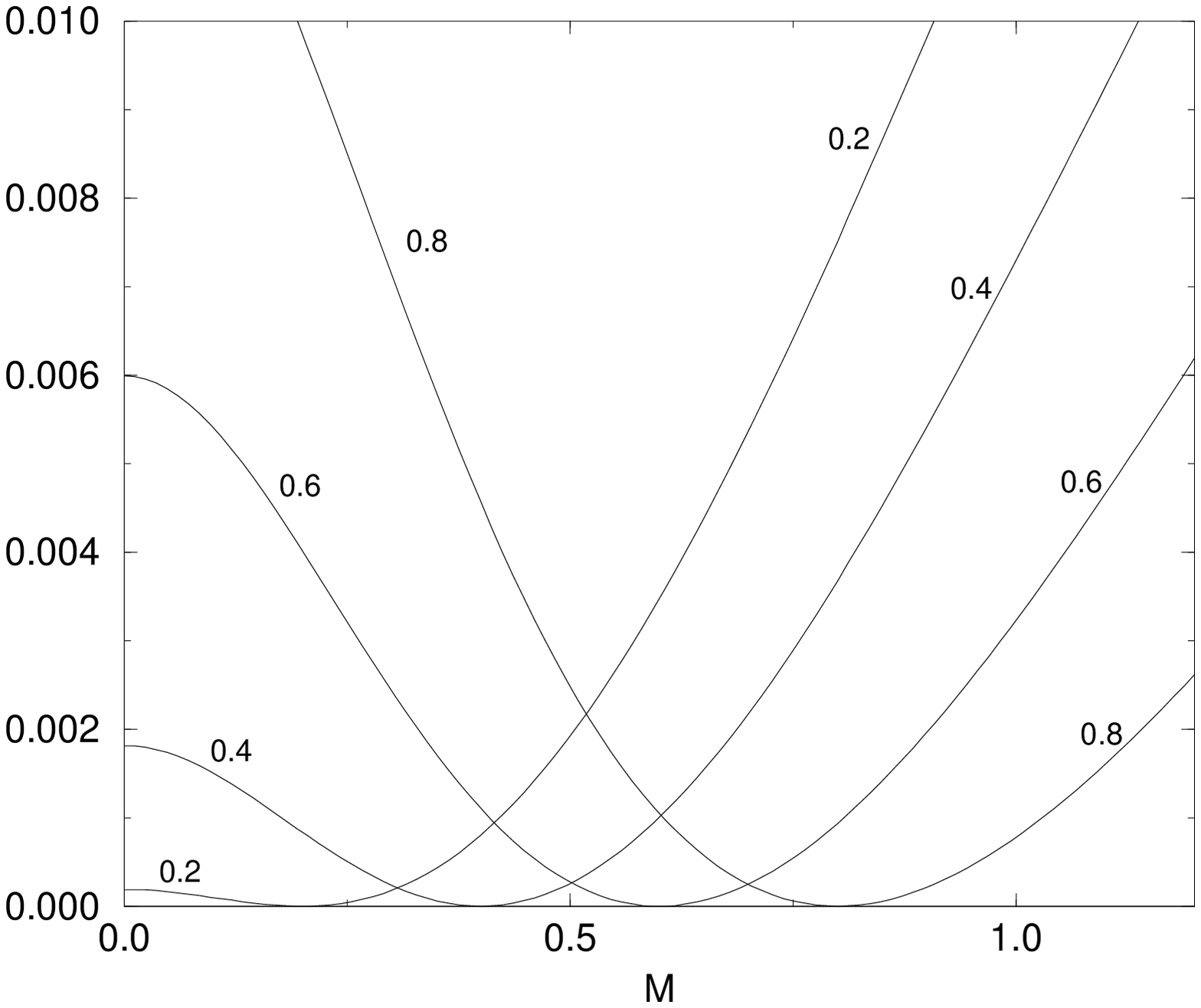}}}
\caption{The effective potential
calculated with a sharp 3-momentum cut-off plotted against $M$. It is
expressed in units of $\Lambda ^{4}$.}
\label{epsfig58}
\end{figure}

Innumerable papers have been published (including some of my own) in which
the meson propagators are derived from the unregulated Nambu Jona-Lasinio
action: 
\begin{equation}
I_{j,m}\left( \varphi \right) =-Tr\ln \left( -i\partial _{\mu }\gamma _{\mu
}+\varphi _{a}\Gamma _{a}\right) -\frac{1}{2}\left( \varphi -m\right) \left(
V-j\right) ^{-1}\left( \varphi -m\right)   \label{notregul}
\end{equation}
The expressions obtained for the propagators are then: 

\begin{eqnarray}
K_{S}^{-1}\left( q\right)  &=&4N_{c}N_{f}\left( \frac{1}{2}\left(
q^{2}+4M^{2}\right) f_{M}\left( q\right) +\frac{m}{M-m}g_{M}\right) 
\label{kspwrong} \\
K_{P}^{-1}\left( q\right)  &=&4N_{c}N_{f}\left( \frac{1}{2}q^{2}f_{M}\left(
q\right) +\frac{m}{M-m}g_{M}\right)   
\end{eqnarray}
where the loop integrals are: 
\begin{equation}
f_{M}\left( q\right) =\frac{1}{\Omega }\sum_{k<\Lambda }\frac{1}{\left(
\left( k-\frac{q}{2}\right) ^{2}+M^{2}\right) \left( \left( k+\frac{q}{2}%
\right) ^{2}+M^{2}\right) }
\end{equation}
and: 
\begin{equation}
g_{M}=\frac{1}{\Omega }\sum_{k<\Lambda }\frac{1}{\left( k-\frac{q}{2}\right)
^{2}+M^{2}}
\end{equation}
The table \ref{table:zapprox} shows the low $q$ behaviour of the $S$ and $P$
inverse propagators in various approximations.\ They are calculated in the
chiral limit.

\begin{table}[tbp] \centering%
%
\begin{tabular}{|c|c|c|c|c|}
\hline
inverse propagators & $K_{P}^{-1}\left( q=0\right) $ & $Z_{\pi }=\left. 
\frac{dK_{P}^{-1}}{dq^{2}}\right| _{q=0}$ & $K_{S}^{-1}\left( q=0\right) $ & 
$\left. \frac{dK_{S}^{-1}}{dq^{2}}\right| _{q=0}$ \\ \hline
regulated action & 0 & 0.0995 & 0.0546 & 0.0592 \\ \hline
regulated $f\left( q\right) $ & 0 & 0.0850 & 0.0544 & 0.0448 \\ \hline
$f\left( q\right) =f\left( 0\right) $ & 0 & 0.0850 & 0.0544 & 0.0850 \\ 
\hline
\end{tabular}
\caption{Three approximations to the inverse $S$ and $P$
propagators, calculated with a sharp 4-momentum cut-off and with $M_{0}/\Lambda =0.4$.\ The first row gives the values obtained from an
regularized action (\ref{ij}).\ The second row gives the values obtained from
a unregularized action and by subsequently regularizing the loop integrals.
The last row gives the results obtained by neglecting the $q$ dependence of
the loop integral $f\left( q\right) $. The inverse quark propagators are
given in units of $\Lambda ^{2}$ and $\frac{dK^{-1}}{dq^{2}}$ is
dimensionless.\label{table:zapprox}}%
\end{table}%
%

\section{An instability of the vacuum.}

The partition function (\ref{saddle}) can also be used to calculate the
effective potential: 
\begin{equation}
\Gamma =W\left( j,m\right) -j\frac{\partial W\left( j,m\right) }{\partial j}%
=W\left( j,m\right) +\frac{1}{2}j\left\langle \left( \bar{\psi}\Gamma
_{a}\psi \right) ^{2}\right\rangle
\end{equation}
As we vary $j$, the squared condensate $\left\langle \left( \bar{\psi}\Gamma
_{a}\psi \right) ^{2}\right\rangle $ changes.\ Thus, when we plot the
effective potential against $j$, we discover how the energy of the system
varies when the system is forced to modify the squared condensate $%
\left\langle \left( \bar{\psi}\Gamma _{a}\psi \right) ^{2}\right\rangle $.
The effective potential has a stationary point at $j=0$, that is, in the
absence of a constraint. If the stationary point of the effective potential
is a minimum, the system is (at least locally) stable against fluctuations
of $\left\langle \left( \bar{\psi}\Gamma _{a}\psi \right) ^{2}\right\rangle $%
.\ If it is an inflection point, it is unstable and we shall indeed find
that this can easily occur when a sharp cut-off is used.

When $j$ is varied, the constituent quark mass $M$ also changes, according
to the gap equation (\ref{gap}).\ One finds that $M$ is a monotonically
increasing function of $j$ so that the effective potential can be plotted
against $M$ equally well. The vacuum constituent quark mass is the mass $%
M_{0}$ obtained with $j=0$. The contribution of each Feynman graph to the
effective potential is stationary at the point $M=M_{0}$ and this is why
plots of the the effective potential against $M$ are nicer to look at than
plots against $j$. The vacuum constituent quark mass $M_{0}$ is a measure of
the interaction strength $V$, to which it is related by the gap equation.
For a given shape of the regulator, the occurrence of an instability depends
on only one parameter, namely $M_{0}/\Lambda $.

Figure \ref{epsfig52} shows the effective potential calculated with a
gaussian cut-off for various values of $M_{0}/\Lambda $.\ The ground state
appears to be stable within the range of reasonable values of $M_{0}/\Lambda 
$.

Figure \ref{epsfig57} shows the effective potential plotted against $M$ when
a sharp cut-off is used. When $M_{0}/\Lambda >0.74$ the ground state
develops an instability with respect to increasing values of $M$. This
instability is not related to the restoration of chiral symmetry and,
indeed, the pion remains a Goldstone boson for all values of $M.$ As shown
on Fig.\ref{epsfig47}, the instability is due to the classical action and
the meson loop contributions do not modify it.

Figure \ref{epsfig58} shows the effective potential calculated with a sharp
3-momentum cut-off.\ No instability appears.\ This provides a clue as to the
cause of the instability which arises when a sharp 4-momentum cut-off is
used.\ Indeed, when a 3-momentum cut-off is used, the Nambu Jona-Lasinio
model defines a time-independent hamiltonian and the 3-momentum cut-off
simply restricts the Hilbert space available to the quarks. This allows a
quantum mechanical interpretation of the results.\ If $H$ is the Nambu
Jona-Lasinio hamiltonian, then the ground state wavefunction $\left|
j\right\rangle $ is calculated with the hamiltonian 
\begin{equation}
\bar{H}_{j}=H-j\int d_{3}x\;\left( \bar{\psi}\Gamma _{a}\psi \right) ^{2}
\end{equation}
containing the constraint proportional to $j$. The effective potential $%
\Gamma $ is then equal to the energy $E\left( j\right) =\left\langle j\left|
H\right| j\right\rangle $ of the system and it displays a stationary point
when $j=0$ or, equivalently, when $M=M_{0}$. The Nambu Jona-Lasinio model,
regularized with a 3-momentum cut-off, has been used in Refs.\cite
{Blaschke94} and \cite{Huefner98} for example.

The use of a 3-momentum cut-off has another important feature. The meson
propagators have only poles on the imaginary axis where they should. When a
4-momentum cut-off is used, unphysical poles appear in the complex energy
plane, as they do when proper-time regularization is used for the quark loop 
\cite{Ripka95}.

The fact that the instability occurs when the model is regularized with a
4-momentum cut-off and not when a 3-momentum cut-off is used, strongly
suggests that the instability is due to the unphysical poles introduced by
the regulator. This conclusion is corroborated by the observation that the
instability also occurs when a gaussian cut-off is used, but at the much
higher values $M_{0}/\Lambda \geq 2.93$ where the cut-off is too small to be
physically meaningful. With a gaussian regulator and in the relevant range
of parameters $0.4<M_{0}/\Lambda <0.8$, one needs to probe the system with
values as high as $M/\Lambda >4$ before it becomes apparent that the energy
is not bounded from below. The instability is an unpleasant feature of
effective theories which use relatively low cut-offs.\ However, the low
value of the cut-off is dictated by the vacuum properties and we need to
learn to work with it. Further details are found in Ref.\cite{Ripka00}.

We conclude from this analysis that it is much safer to use a soft
regulator, such as a gaussian, than a sharp cut-off.

\bibliographystyle{unsrt}
\bibliography{njl}

\begin{thebibliography}{10}

\bibitem{Kleinert99}
H.~Kleinert and B.~Van~Den Boosche.
\newblock {\em Phys.Lett. B474}, page 336, 2000.

\bibitem{Ripka00}
G.Ripka.
\newblock Quantum fluctuations of the quark condensate.
\newblock hep-ph/0003201, 2000.

\bibitem{Zinn89}
J.Zinn-Justin.
\newblock {\em Quantum Field Theory and Critical Phenomena}.
\newblock Clarendon Press, Oxford, 1989.

\bibitem{Diakonov96}
C.~Weiss D.I.~Diakonov, M.V.~Polyakov.
\newblock Hadronic matrix elements of gluon operators in the instanton vacuum.
\newblock {\em Nucl. Phys. B461}, page 539, 1996.

\bibitem{Birse95}
R.D.Bowler and M.C.Birse.
\newblock {\em Nucl.Phys. A582}, page 655, 1995.

\bibitem{Birse98}
R.S.Plant and M.C.Birse.
\newblock {\em Nucl.Phys. A628}, page 607, 1998.

\bibitem{Ripka98}
W.Broniowski B.Golli and G.Ripka.
\newblock {\em Phys.Lett. B437}, page~24, 1998.

\bibitem{Broniowski99}
Wojciech Broniowski.
\newblock Mesons in non-local chiral quark models.
\newblock hep-ph/9911204, 1999.

\bibitem{Broniowski00}
B.Szczerebinska and W.Broniowski.
\newblock {\em Acta Polonica 31}, page 835, 2000.

\bibitem{Ripka97}
Georges Ripka.
\newblock {\em Quarks Bound by Chiral Fields}.
\newblock Oxford University Press, Oxford, 1997.

\bibitem{Musakhanov99}
M.Musakhanov.
\newblock {\em Europ.Phys.Journal C9}, page 235, 1999.

\bibitem{Shuryak82}
E.Shuryak.
\newblock {\em Nucl.Phys. B203}, pages 93,116,140, 1982.

\bibitem{Diakonov86}
D.I.Diakonov and V.Y.Petrov.
\newblock {\em Nucl.Phys. B272}, page 457, 1986.

\bibitem{Blaschke94}
D.Blaschke S.Schmidt and Y.Kalinovsky.
\newblock {\em Phys.Rev. C50}, page 435, 1994.

\bibitem{Huefner98}
S.P.~Klevansky Y.B.~He, J.~H'fner and P.~Rehberg.
\newblock {\em Nucl. Phys. A630}, page 719, 1998.

\bibitem{Ripka95}
E.N.Nikolov W.Broniowski, G.Ripka and K.Goeke.
\newblock {\em Zeit.Phys. A354}, page 421, 1996.

\end{thebibliography}

\end{document}